\documentclass[10pt]{article}

\usepackage{amsmath}
\usepackage{amssymb}

\usepackage{graphicx}
\usepackage{subfig}

\usepackage{cite}

\usepackage{color} 

\usepackage[labelfont=bf,labelsep=period,justification=raggedright]{caption}

\makeatletter
\renewcommand{\@biblabel}[1]{\quad#1.}
\makeatother

\date{}

\begin{document}

\begin{flushleft}
{\Large
\textbf{The role of neighbours selection on cohesion and order of swarms.}
}
\\
Angelo M. Calv\~ao$^{1}$, 
Edgardo Brigatti$^{2,\ast}$, 
\\
\bf{1} Instituto de F\'{\i}sica, Universidade Federal Fluminense, Campus da Praia Vermelha, Niter\'oi, RJ, Brazil
\\
\bf{2} Instituto de F\'{\i}sica, Universidade Federal do Rio de Janeiro, Cidade Universitária, Rio de Janeiro, RJ, Brazil
\\
$\ast$ E-mail: edgardo@if.ufrj.br
\end{flushleft}

\section*{Abstract}

We introduce a multi-agent model for exploring 
how selection of neighbours determines
some aspects of order and cohesion in swarms.
The model algorithm states that every 
agents' motion seeks for an optimal 
distance from the nearest topological 
neighbour encompassed in a limited attention field.
Despite the great simplicity of the implementation, 
varying the amplitude of the attention landscape,
swarms pass from cohesive and regular 
structures towards fragmented and irregular
configurations.
Interestingly, this movement rule is an ideal candidate for 
implementing the selfish herd hypothesis which
explains aggregation of alarmed group of social animals.

\section*{Introduction}
\label{introduction}

Collective group motion is an important biological phenomenon that has received much empirical and theoretical attention from investigators in disciplines as varied as computer science  \cite{boids},  biology \cite{okubo, biolog0, biolog1, biolog2, couzin}  and physics \cite{vicsek1,gregoire,ginelli}. 
Generally speaking two types of mechanisms are considered:
an aligning interaction, and attraction/repulsion between individuals \cite{gautrais}.
Usually, the first one is the responsible for 
the emergence of 
polarised groups \cite{vicsek2} ,  
the second one for maintaining cohesion
with a rather homogeneous density which can correspond to
a certain level of regularity in the spacial distribution. 
In this work we will focus only on the second.

Cohesion and regularity in the spacial distribution 
can generate many biological advantages. 

First, we can list energetic benefits.
A classical example is given by
flocks of birds which align themselves 
in ``V" formations  \cite{geese}. 
In this situation, 
individuals seek an
optimal mutual position 
which generates a regular structure in the distribution of inter-individual distances.
Other examples of this behaviour can be found in
the core of big herds of migrating mammals \cite{biolog1},
and in flocks of surf scoters 
moving on the water surface.
Very detailed observations showed that individuals seek a target density
searching for an ideal distance from the other components of the group,
generating well-defined spatial structures \cite{scoters}.
As inferred by a recent field study on  mosquitofish \cite{mosquitofish},
on the one hand, animals move away from individuals that intrude their personal space,
defining a stress zone, 
while, on the other hand,
they are attracted to individuals at a significant distance.
This behaviour allows to maintain integrity as a group  while 
decreasing the frequency of abrupt accelerations or decelerations. 

Second, aggregation can improve
reproductive success \cite{burger}.
An astonishing case is displayed by the males 
of a cichlid fish which mark their breeding 
territories digging pits. They design hexagonally shaped 
territories that reach an impressive high degree of order \cite{ciclids}.

Finally, a particularly important
benefit is the reduction of predation risk.
The study of this aspect 
generated an interesting conjecture for explaining 
aggregation of individuals \cite{hamilton}.
This 
idea, known as the  ``selfish herd hypothesis",
suggests that cohesive swarms are generated
because individuals move toward one another 
for minimising their own predation risk.
Some recent field observations presented evidences 
of this behaviour analysing the movements of sand fiddler crabs \cite{viscido1}
and seals \cite{devos}.
This last study revealed that simple movement
rules are used to reduce predation risk.
Effectively, a successful  implementation of this hypothesis 
require an individual movement rule 
as simple as possible.
This is a necessary condition for many species, 
across different taxa, being able to follow it 
and for enabling natural selection to fix it \cite{reluga}. 

Unfortunately, simple rules tested by computer simulations 
seems to fail to produce aggregation \cite{morton, viscido2}.
Specifically, the implementation of the Hamilton's 
algorithm, where the focal individual moves 
towards the nearest neighbour, does not result in compact
and dense groups and fragmentation into multiple 
sub-groups of a few individuals occurs. 
Discovery an elementary movement rule
which can produce compact aggregation 
is still an unsolved problem which
has been denominated the ``dilemma of the selfish herd" \cite{viscido2}.

In the following we introduce a very simple algorithm which 
seems to find a solution to this dilemma:
we consider an interaction where individuals move toward
the nearest neighbour encompassed in a limited attention field.
We stress how this rule
is more natural than 
asking individuals to make their decisions averaging
over the influence of many neighbours \cite{biolog0, gregoire, viscido2}.
In fact, in our case, cohesion is determined by the number of interactions
and not by abstract parameters which control the
interaction like in a molecular type force.
It is difficult to imagine that organisms behave like 
particles in a physical system, where interaction is
mediated by a potential which directly sums up for all
the neighbours.
Individuals must undertake efficient decision-making, 
instead of relying on the 
weighted sum of a large number of neighbours.
For this reason, they choose only the
neighbours relevant for their purpose and they face a
natural limitation over the information they can handle.
These limits are not just 
shaped by the physiology of vision or the visual system response;
perceptual and cognitive effects should be the most relevant ones.
Among them, the physiology of attention should be  
particularly important. When we are confronted 
with a large number of items, we withdraw from most and we focus our 
attention on just a few. The visual system's neurones are responsive 
to what neighbours we are interested in \cite{attention1, attention2, attention3}.  
We simplify these considerations imagining 
that individuals make decisions based 
on an angular landscape defined by their attention.
In this attention field, the organism fixes on the 
closest neighbour and reaches a preferred distance 
in relation to it.
The fact that this rule is based on a topological 
interaction and depends 
on perceptual limits
are realistic aspects outlined by field observations  \cite{cavagna,ballerini}.
Moreover, a recent study,  which explicitly determined 
the interaction rules in fish groups, identified the
single nearest neighbour interaction, 
applied with the aim of active regulating the 
distance between pair of animals,
as the principal mechanism for collective motion  \cite{mosquitofish}.

Our simple algorithm is capable of 
exhibiting a rich 
behaviour characterised by different phases.
Changing the amplitude of the attention 
landscape swarms pass from cohesive and regular 
structures towards fragmented and irregular
configurations.
Different levels of angular and positional order
are spanned and described.
In particular, it becomes evident that only 
dealing with a reduced portion of the attention field
can generate a cohesive and ordered swarm.
In this regime the algorithm is an ideal candidate for implementing 
the selfish herd hypothesis.\\

The paper is organised as follows. In the following section we introduce the model. 
In Section~\textit{Analysis of the model behaviour}, an in depth analysis of the algorithm is performed, 
giving a clear overview of its performance in different regimes.
In Section~\textit{Application to the selfish herd problem}, we apply our movement rule to the selfish herd problem and we compare 
qualitatively our results with field observations. Conclusions are reported in
the last section.

\section*{The model}
\label{sec:the_model}

We consider a system composed by $P$ agents which move continuously
on a square of linear size L and which stop if they reach the boundary.
In any case, the definition of the specific 
form of the boundary condition is not relevant, because, in practice, 
individuals never approach the boundary.
The time unit $\tau$ is the time interval between two updatings of 
the positions of all the agents. 
In most of our simulations the initial conditions correspond
to $91$ individuals uniformly and randomly distributed on the plane
with a density $\delta=30$. 
The stress zone radius $D$ is fixed to 0.1 length unit.

Movements are determined asynchronously.
An agent $i$, with position $\mathbf{x}$, is randomly selected and a 
gazing direction  is assigned by a random number chosen with a uniform 
probability from the interval $[0, 2\pi]$.
This probability distribution is the simplest
hypothesis for a general implementation of our algorithm
and it can be supported by some observational results \cite{scoters}.
The gazing direction is the bisector of the angle $\Theta$,
which defines the attention field where the nearest neighbour is  sought.
It is important to implement this search by using a fast algorithm. 
We employed the Computational Geometry Algorithms Library, using the
2D Range and Neighbor Search \cite{algorithm}.

Given the position $\mathbf{y}$ of this neighbour, 
if $|\mathbf{x_\tau}-\mathbf{y}|>D$, $i$ moves in the direction 
of its neighbour a step $\epsilon$: 
$\mathbf{x_{\tau+1}}=\mathbf{x_\tau}+\epsilon \mathbf{v}$,  
where $\mathbf{v}= \frac{\mathbf{y}-\mathbf{ x_\tau}} {|\mathbf{y}-\mathbf{x_\tau}|}$.
If this movement 
results in the stress zone invasion, 
$i$ stops along the direction $\mathbf{v}$ 
at a distance $D$ from its neighbour. 

\begin{figure}[ht]
\centering
\includegraphics[scale=0.8]{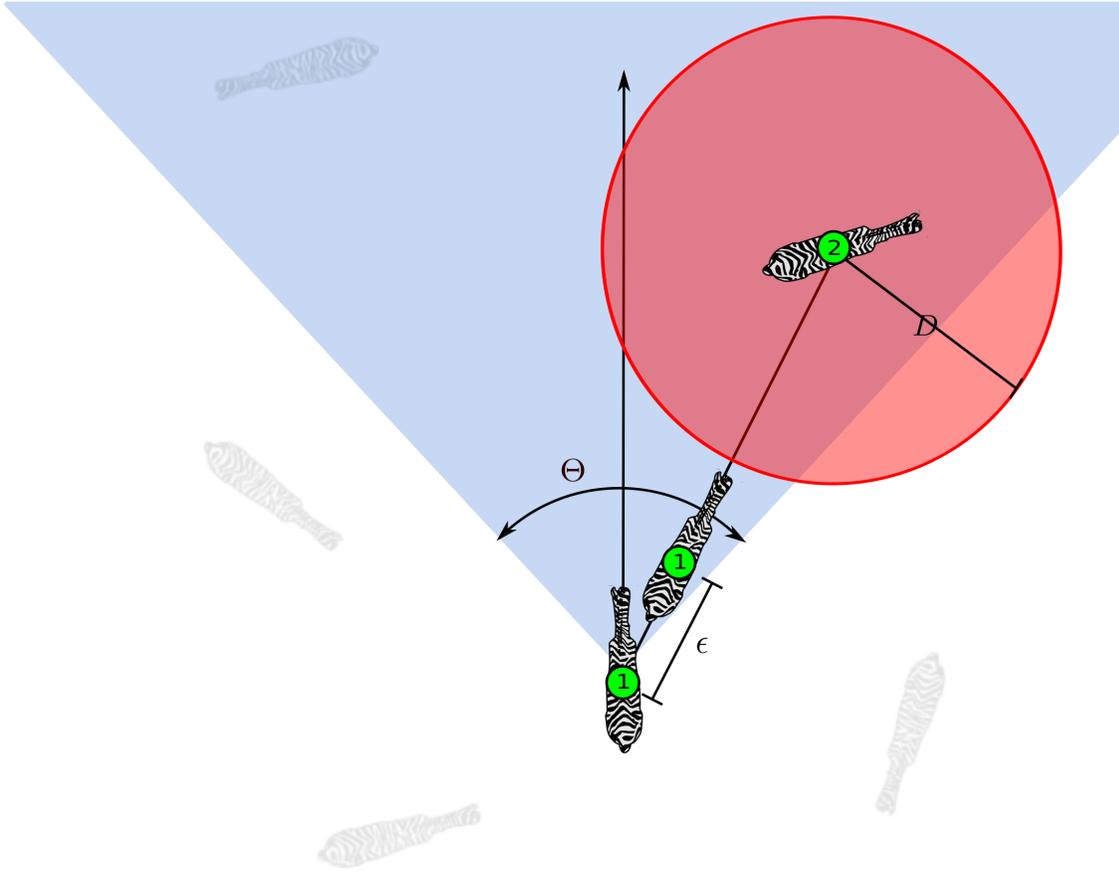}
\caption{{\bf An example of an agent's movement.}
Individual 1, moves with speed $\epsilon$
towards individual 2 which is
the nearest neighbour encompassed by
the attention field, the blue region characterised by the angle
$\Theta$.
The stress zone of individual 2 is the red disk of radius $D$. 
}
\label{fig:steps}
\end{figure}

If $|\mathbf{x_\tau}-\mathbf{y}|<D$, the movement is: 
$\mathbf{x_{\tau+1}}=\mathbf{x_\tau}-\epsilon \mathbf{v}$.
If $i$ gets farther than $D$ from its neighbour, $i$ stops at a 
distance equal to $D$. Figure \ref{fig:steps} gives an example
of these simple rules. 

This algorithm considers only the relative movements 
of the individuals.
Obviously, for describing a directed moving swarm,
a superimposed common collective velocity can be  added
but it is not relevant for our analysis.
 
\begin{table}[ht]
    \centering
{\small
    \begin{tabular}{|c|p{70px}|c|p{120px}|c|}
\hline
Parameter & Description & Typical value & Biological interpretation  
& Typical unit \\
\hline
\hline
D & stress zone radius & 0.1 & average inter-individual  distance     &  10 $\times$ m \\
\hline
L &  linear size of the box & 1.74 & linear size of the observation region & 10 $\times$ m \\
\hline
$\epsilon$ & speed 
& 0.0112 & average velocity & 10 $\times$ m/s \\
\hline
P & group size & 91 & group size & individuals \\
\hline
$\Theta$ & amplitude of the attention field & $70^o - 360^o$  & average attention field & degree \\
\hline
\end{tabular}
}
\caption{
    { \bf Parameters of the model. }
The typical values are used in the  simulation trials of section
\textit{Analysis of the model behaviour} and, in this case, 
$D$, $L$ and $\epsilon$ are expressed in length units.
The column ``typical unit" refers to the simulation of a swarm of crabs
\cite{viscido1} as described in section \textit{Application to the selfish herd problem}}
\label{table:params}
\end{table}

Simulation parameters are summarised in Table~\ref{table:params}.
Our C++ computer code is available upon request.

\section*{Results}
\label{sec:results}

\subsection*{Analysis of the model behaviour}
\label{sec:model_analysis}

The principal purpose of this section 
is to quantify, 
with respect to the amplitude of the attention field, 
which degrees of cohesion and order our algorithm is able to produce.
For this reason, the algorithm operates until a quasi-stationary 
state is reached.
 We consider this state because a neat and clear 
analysis can be performed
and not because we are interested in static aggregation.
In fact, the active configurations displayed along 
the dynamics are qualitatively identical to the final 
quasi-stationary state.
We affirm that our system entered this state if, 
during $30$ time steps, no changes in agents' positions is recorded.
This state can correspond to an effective absorbing state, where 
all the agents' distances with all their topological neighbours are equal to $D$.
Otherwise, it is possible that some movements, even if improbable,
would be still feasible. 
In this last case, the agents' distances from their 
first metric neighbour are equal to $D$.

\begin{figure}[ht]
\begin{center}
        \input{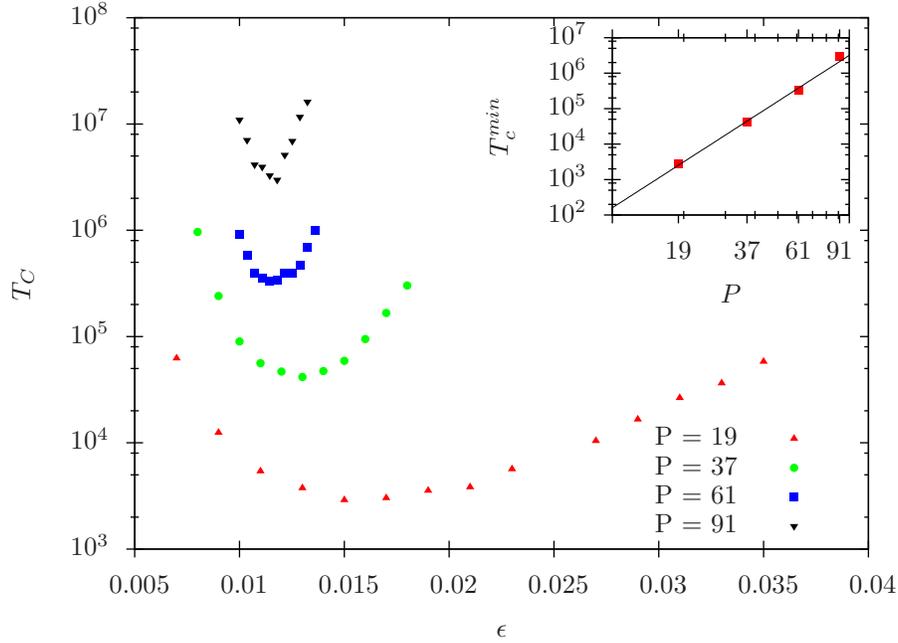}
\end{center}
\caption{ 
    {\bf The convergence time as a  function of the speed $\epsilon$,
        for different populations $P$ ($\Theta=75^{\circ}$, $\delta=15$).}
Data points represent averages 
taken over $100$ different simulations, where
the individual initial distribution is different.
In the inset: $T_C^{min}$ as a function of $P$.
The continuous line is the fitted relation:
 $T_C^{min} \propto P^{4.4}$.
}
\label{fig_timeEp}
\end{figure}

First, we investigate the behaviour of the convergence time 
for reaching a quasi-stationary state ($T_C$). 
Interestingly, the $T_C$ value 
strongly depends 
on $\epsilon$ and an optimal $\epsilon_o$ value exists, 
for which $T_C$ is minimal (Figure~\ref{fig_timeEp}).
In addition, varying the population size, 
the optimal $\epsilon_o$ value slightly changes along with the value of $P$.
Since we start all simulations with the same density, different
convergence times are not 
caused by different density values, but by 
a collective effect in the ordering procedure.
In Figure~\ref{fig_timeEp} we display the 
convergence times for different 
population size when the $\epsilon_o$
value is chosen ($T_C^{min}$).
As can be appreciate, $T_C^{min}\propto P^{4.4}$.\\

Now we focus on the description of the dynamics of the model.
We introduce two order parameters which 
are able to capture the degree of order reached by the 
swarm. We characterise the degree of positional orientational order 
of a given configuration defining, for each organism $j$  \cite{nelson, strandburg}:
$\psi_j= \frac{1}{N_j}\sum^{N_j}_{k=1} \exp(i6\phi_{jk}),$
where $N_j$ is the number of topological neighbours of 
individual $j$, 
which are the organisms whose Voronoi polygons share an edge with 
individual $j$.
Finally, $k$ is the index of the neighbours and 
$\phi_{jk}$ is the angle relative to the bond between 
$j$ and $k$ and an arbitrary fixed reference axis.
The factor of $6$ is introduced for detecting 
perfect sixfold ordered structures.
A positional orientational order parameter is given by the norm of the 
average of $\psi_j$ over all the organisms $j$:

\begin{equation}
\psi_6= \frac{1}{P}\left| \sum^{P}_{j=1} \psi_j \right|.
\label{eq:psi_6}
\end{equation}

\begin{figure}[ht]
\begin{center}
    \input{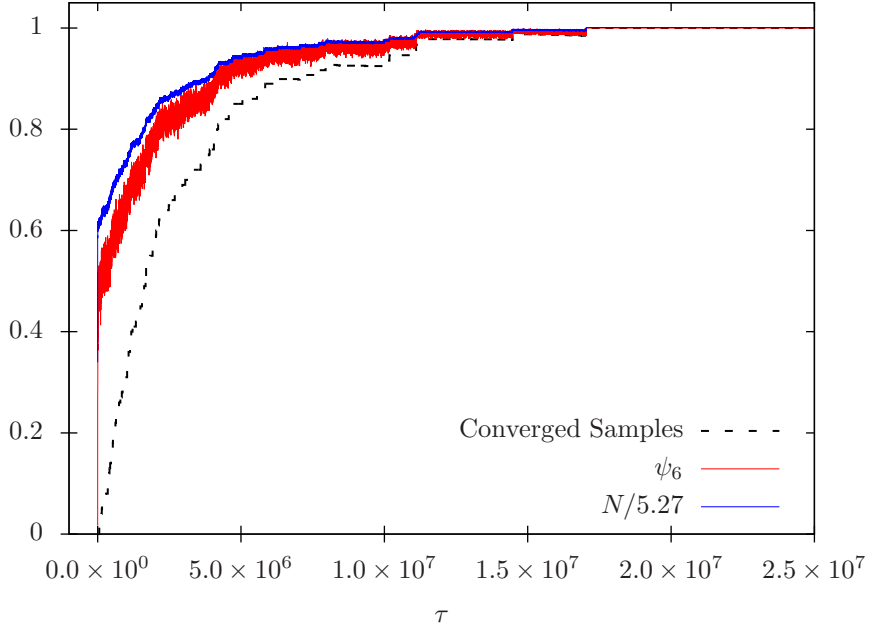}
\end{center}
\caption{ {\bf Time evolution of the order parameters $N$ and $\psi_6$
averaged over 100 simulations.}
The dashed line is the convergence probability of the ensemble of simulations.
$P=91$, $\Theta=75^{\circ}$, $\epsilon=0.0112$, 
and $\delta=30$.
}
\label{fig_Dynamics}
\end{figure}

Translational order can be investigated by looking at 
the sum of the number of individuals 
contained in a circle of radius $D+0.001$ around a given individual.
We obtain the translational ordering parameter $N$ averaging this quantity 
over all the individuals.
Figure~\ref{fig_Dynamics} shows the time evolution of these two order parameters
for typical values which generate a final state characterised by a perfect 
sixfold ordered structure (absorbing state).\\

In the following we analyse the cohesion and order of the swarm
varying the attention field angle $\Theta$.
For this purpose, we look at 
the quasi-stationary states,
which clearly reveal
the differences in the ordering ability of the algorithm
for different values of $\Theta$ (see Figure \ref{fig_Figpheno}).
This fact forces us to run very long simulations
where almost the entire computational time is 
lost in searching for the nearest neighbour inside a given attention field.

\begin{figure}
\begin{center}
       \subfloat[$\Theta<60^o$  Connected and disordered]{
        \scalebox{0.9}{\includegraphics{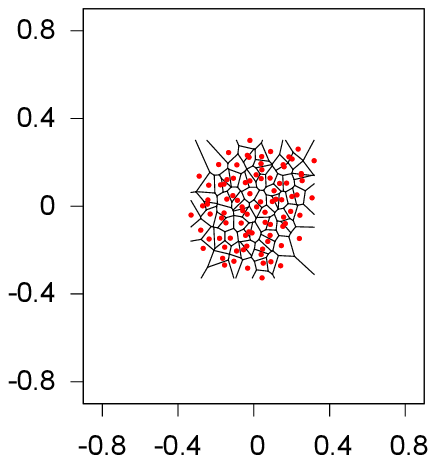}}
        \scalebox{0.9}{\includegraphics{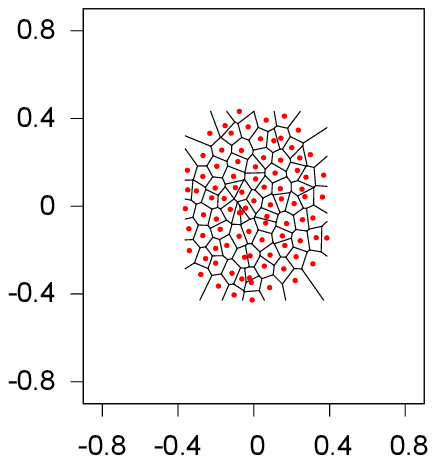}}
        \scalebox{0.9}{\includegraphics{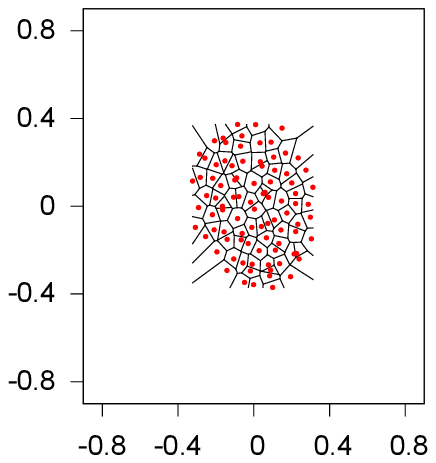}}
    }

    \subfloat[$60 < \Theta < 210^o$  Connected and ordered ]{ 
        \scalebox{0.9}{\includegraphics{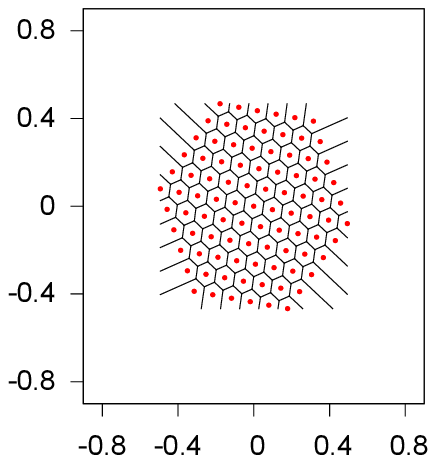}}
        \scalebox{0.9}{\includegraphics{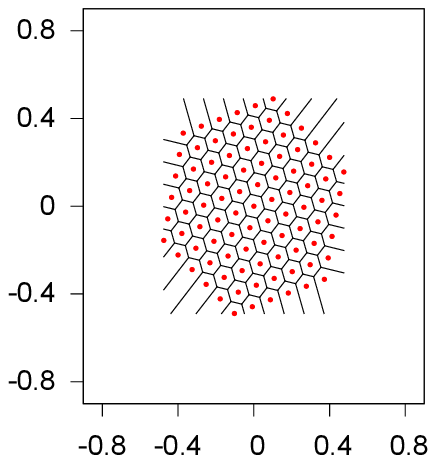}}
        \scalebox{0.9}{\includegraphics{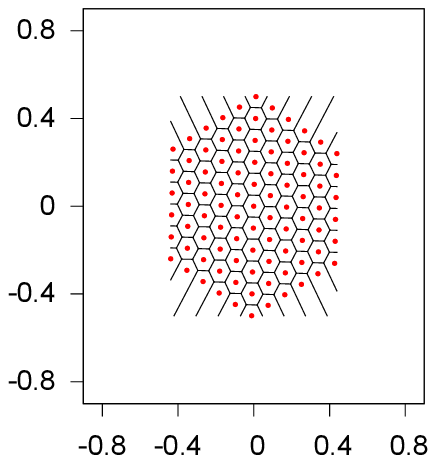}}
    }

    \subfloat[$\Theta = 210^o$  Defects appear]{
        \scalebox{0.9}{\includegraphics{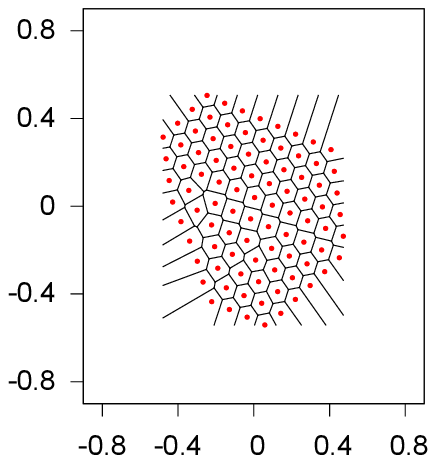}}
        \scalebox{0.9}{\includegraphics{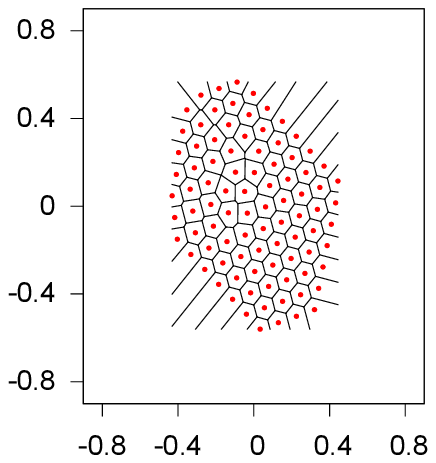}}
        \scalebox{0.9}{\includegraphics{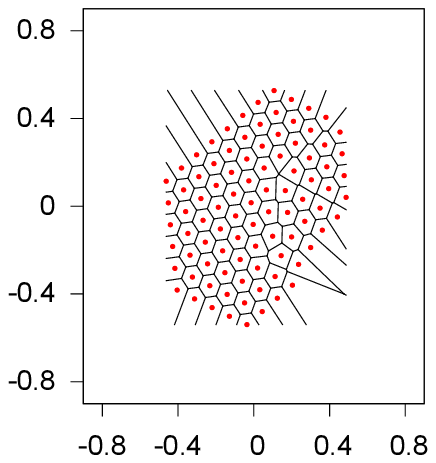}}
    }
\end{center}
\caption{ {\bf Examples of some characteristic configurations of the
    quasi-stationary states for different $\Theta$ values ($P=91$,
$\epsilon=0.0112$ and $\delta=30$).}
The red dots represent the agents' position and the lines depict the Voronoi tessellation.
}
\label{fig_Figpheno}
\end{figure}

\begin{figure}
    \ContinuedFloat 
\begin{center}
    \subfloat[$\Theta = 225^o$ Holes appear]{
        \scalebox{0.9}{\includegraphics{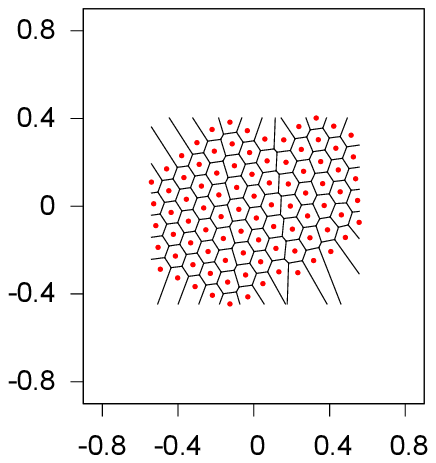}}
        \scalebox{0.9}{\includegraphics{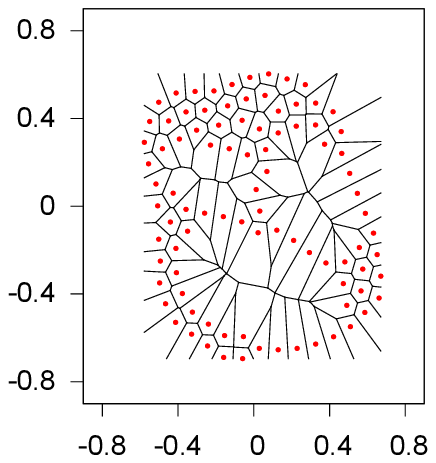}}
        \scalebox{0.9}{\includegraphics{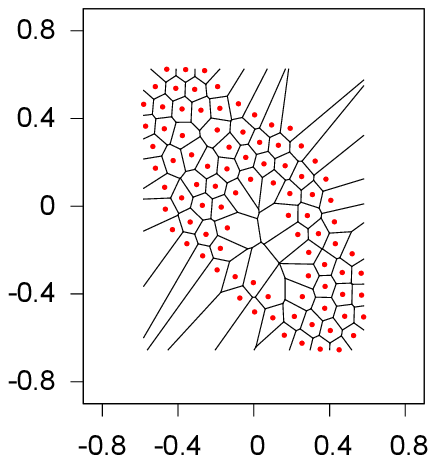}}
    }

    \subfloat[$\Theta = 270^o$ Connected filaments]{ 
        \scalebox{0.9}{\includegraphics{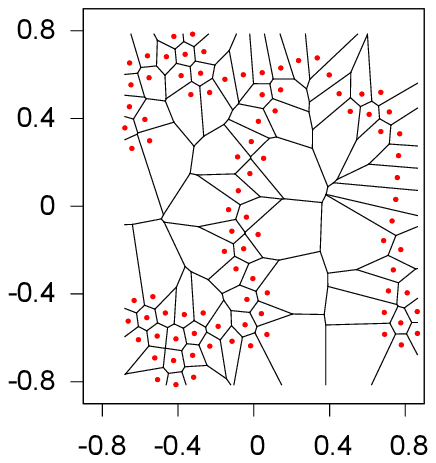}}
        \scalebox{0.9}{\includegraphics{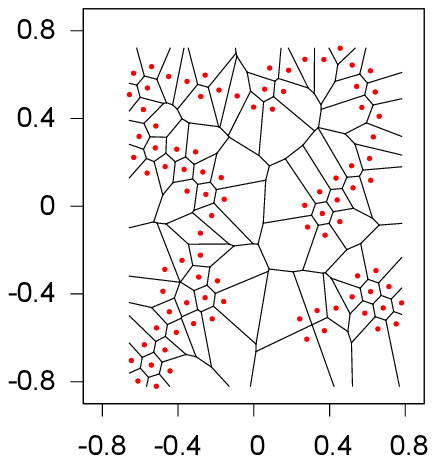}}
        \scalebox{0.9}{\includegraphics{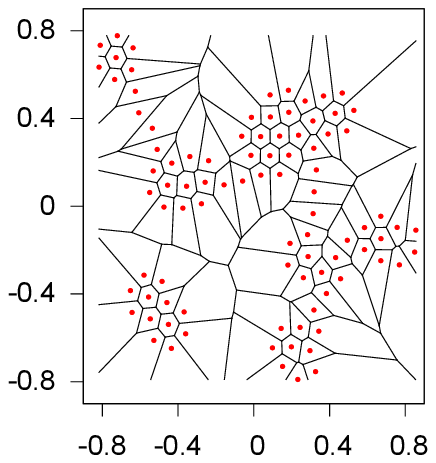}}
    }

    \subfloat[$\Theta = 330^o$ Fragmented]{ 
        \scalebox{0.9}{\includegraphics{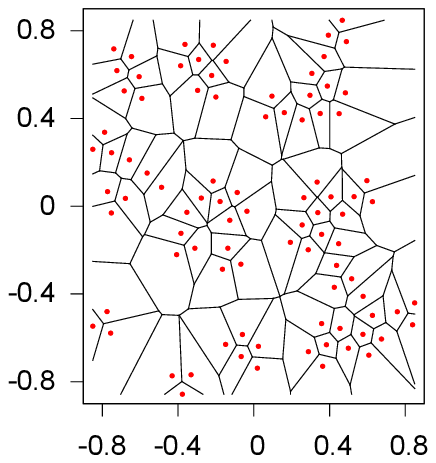}}
        \scalebox{0.9}{\includegraphics{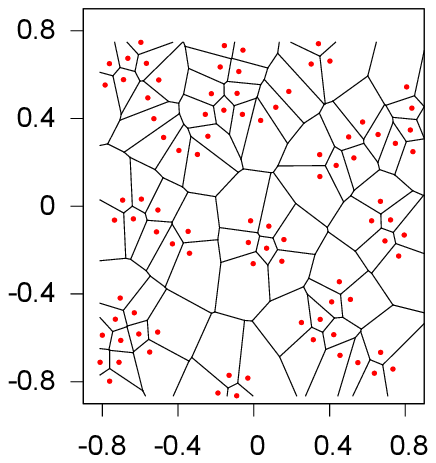}}
        \scalebox{0.9}{\includegraphics{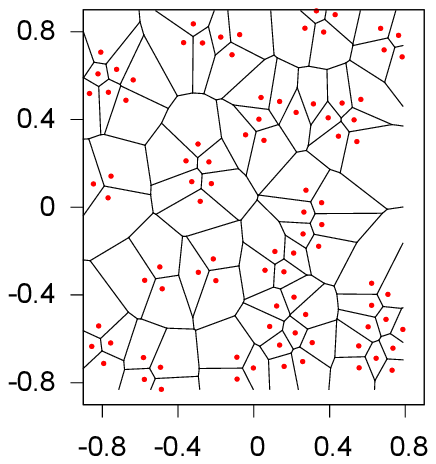}}
    }
\end{center}
\caption{ {\bf Examples of some characteristic configurations of the
    quasi-stationary states for different $\Theta$ values ($P=91$,
$\epsilon=0.0112$ and $\delta=30$).}
The red dots represent the agents' position and the lines depict the Voronoi tessellation.
}
\label{fig_Figpheno}
\end{figure}

For small $\Theta$ values the swarms maintain a high level of cohesion,
but a totally disordered configuration.
The groups reach a high density and individuals do not respect 
the stress zone.  The quasi-stationary states are not attained and 
individuals continue to change their relative positions.
For $\Theta$ values higher than $60\,^{\circ}$, the interaction
arranges the swarm in the densest way compatible with a 
pairwise distance equal to $D$. In fact, 
all the organisms are located on the vertices
of equilateral triangles, which tile the plane along the 
six-fold symmetric triangular lattice.
This phase can be easily detected looking at the value of 
$\psi_6$ which is equal to 1.
Raising the value of $\Theta$ 
around  $220\,^{\circ}$ lattice defects begin to appear 
in the form of individuals with a number of neighbours different from 6 
(generally 5 and 7).
The disordered phase (liquid) emerge for  $220\,^{\circ}<\Theta<300\,^{\circ}$.
In this interval, increasing the value of $\Theta$
generates holes in the swarm structure. 
These ruptures can significantly grow generating
linear structures of particles, which result in sub-swarms
connected by filaments.
Finally, for larger  $\Theta$ values ($\Theta>300\,^{\circ}$)
cohesion is lost and isolated clusters of organisms appear.

\begin{figure}[ht]
\begin{center}
    \input{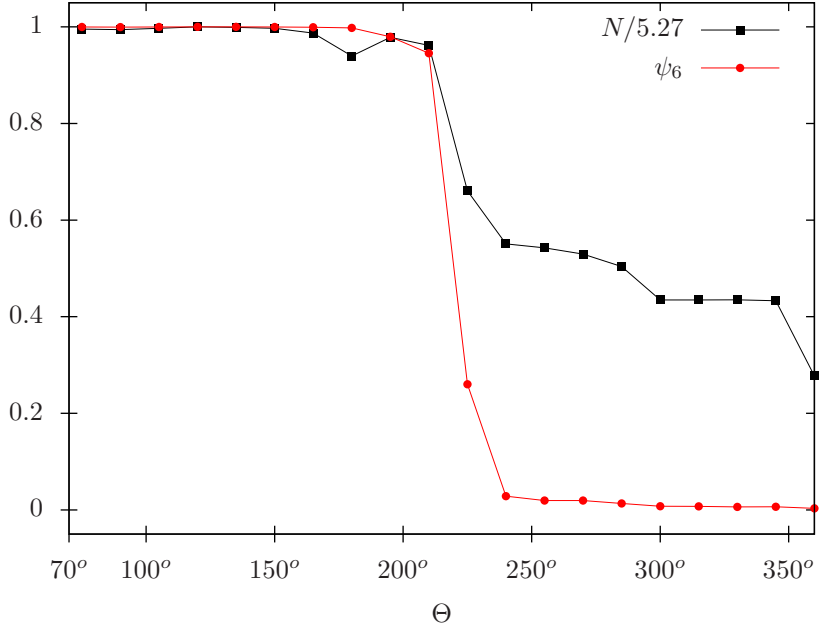}
\end{center}
\caption{{\bf Order parameters $N$ and $\psi_6$ for different $\Theta$.}
Data are averaged over 100 simulations
 when the system reaches the quasi-stationary state
($P=91$, $\epsilon=0.0112$, 
and $\delta=30$).
}
\label{fig_TransPheno1}
\end{figure}

\begin{figure}[ht]
\begin{center}
    \input{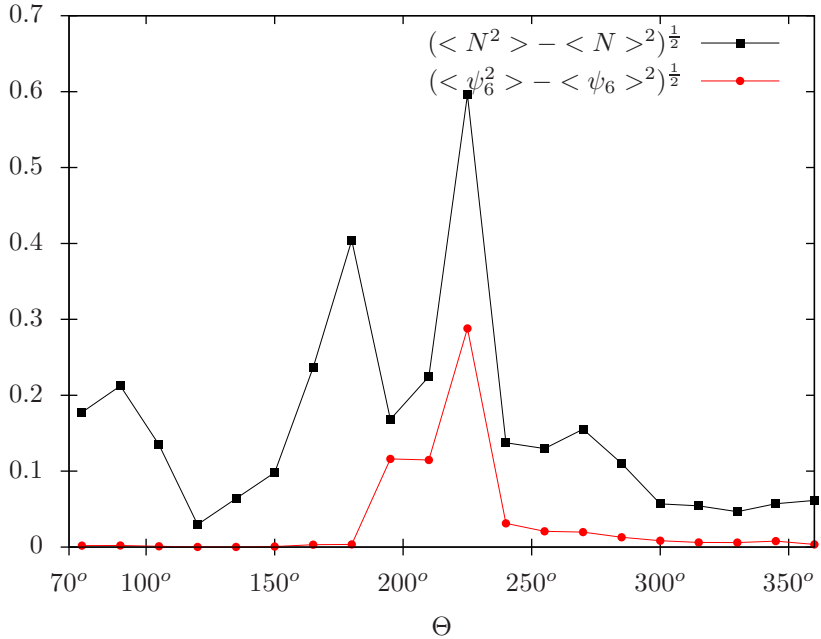}
\end{center}
\caption{ 
{\bf   The standard deviations of $\psi_6$ and $N$ for different $\Theta$.}
Data are averaged over 100 simulations
 when the system reaches the quasi-stationary state
($P=91$, $\epsilon=0.0112$, 
and $\delta=30$).
}
\label{fig_TransPheno2}
\end{figure}

An abrupt variation in the $\psi_6$ and $N$ values signal these 
transitions (Figure~\ref{fig_TransPheno1}). 
The transition between the ordered and the disordered
phase ($\Theta=220\,^{\circ}$) is clearly detected by the 
drops in the $\psi_6$ and $N$ values.
Moreover, in analogy with equilibrium phase transitions, 
the fluctuations of the order parameters increase on approaching this critical value. 
This is shown in  Figure~\ref{fig_TransPheno2} by the standard deviations
of $\psi_6$ and $N$, which exhibit a sharp peak in correspondence of 
$\Theta=220\,^{\circ}$.
The second transition between the cohesive and the 
fragmented phase ($\Theta=300\,^{\circ}$) is evidenced
by a decrease in the $N$ parameter, which reaches
a plateau value close to 2. 

To sum up, we can highlight four different phases:
a first disordered, highly dense, connected phase ($\Theta<60\,^{\circ}$),
a crystal-like phase, with  hexagonal patterns
($60\,^{\circ}<\Theta<220\,^{\circ}$),
a low ordered phase with the presence of 
holes and ruptures and, finally, for
$\Theta>300\,^{\circ}$, a fragmented
swarm, where the initial group splits 
in different clusters and cohesion is lost. 
An interesting dependence of this behaviour 
on the value of the initial density was found and 
will be published any time soon.\\

A natural interest exists for introducing
a noise source in our ordering algorithm.
Noise is an obvious element presents in real systems and
configurations obtained with the presence of noise can have a stronger relation
with real-life situations.
In addition, from a theoretical point of view, it is interesting to state if 
this new ingredient can generate some type of order-disorder
phase transition.
For this reasons, we introduce a 
noise in the evaluation of the direction of the displacement of each organism.
With this new rule, an agent, after having determined 
the vector $\mathbf{v}$, changes the orientation
of its movement by a random angle chosen
from the interval $[-\eta\pi,\eta\pi]$ with a uniform probability.
This means that the final direction of the movement is obtained 
after rotating the original direction $\mathbf{v}$ with a random angle
and  $\eta$ is the parameter which controls the noise strength.
The speed continues to be equal to $\epsilon$, with the same restriction
of the deterministic case when the ideal distance $D$ is crossed.

\begin{figure}[ht]
\begin{center}
    \input{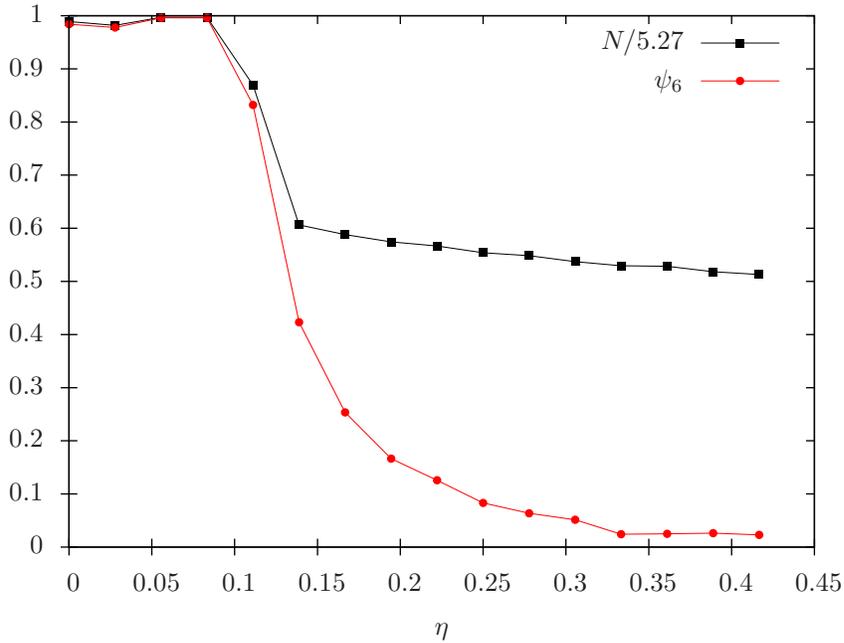}
\end{center}
\caption{ 
    {\bf $N$ and $\psi_6$ as a function of the noise strength $\eta$
    after $12\times10^6$ iterations.}
Data are averaged over 100 simulations 
($P=91$, $\Theta=75^{\circ}$, $\epsilon=0.0112$ and $\delta=30$).
}
\label{fig_Ntransition}
\end{figure}

We fix $\Theta =75\,^{\circ}$, which, for $\eta=0$, generates
the ordered six-fold configurations.
As can be seen in Figure~\ref{fig_Ntransition}
an order-disorder transition emerges, where the 
disordered phase is characterised by $\psi_6\approx0$.
We can clearly appreciate the abrupt appearance of a 
spatial order for a critical value of the noise,
where a collective motion is attained for sufficiently 
low levels of noise.

\subsection*{Application to the selfish herd problem}
\label{sec:application_selfish_herd}

\begin{figure}[ht]
\begin{center}
    \input{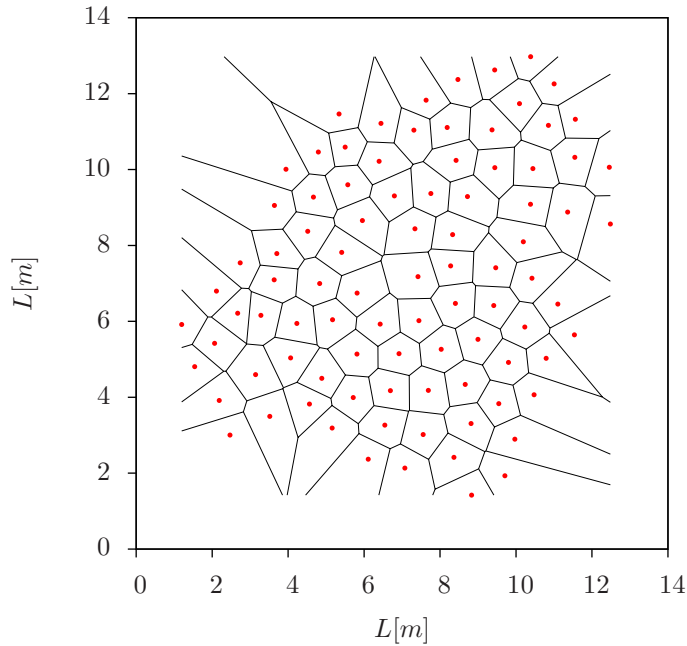}
\end{center}
\caption{ 
    {\bf An active configuration obtained using the parameter 
$P=90$, $\Theta=160^{\circ}$, $\eta=0.55$, $D=1~m$, 
$\epsilon=0.2~m/s$,  and $L=14~m$.}
After a rapid transient, during which the group density increases, 
we obtain a configuration qualitatively similar to the field observations
of reference \cite{viscido1}.
}
\label{fig_compare1}
\end{figure}

\begin{figure}[ht]
\begin{center}
    \input{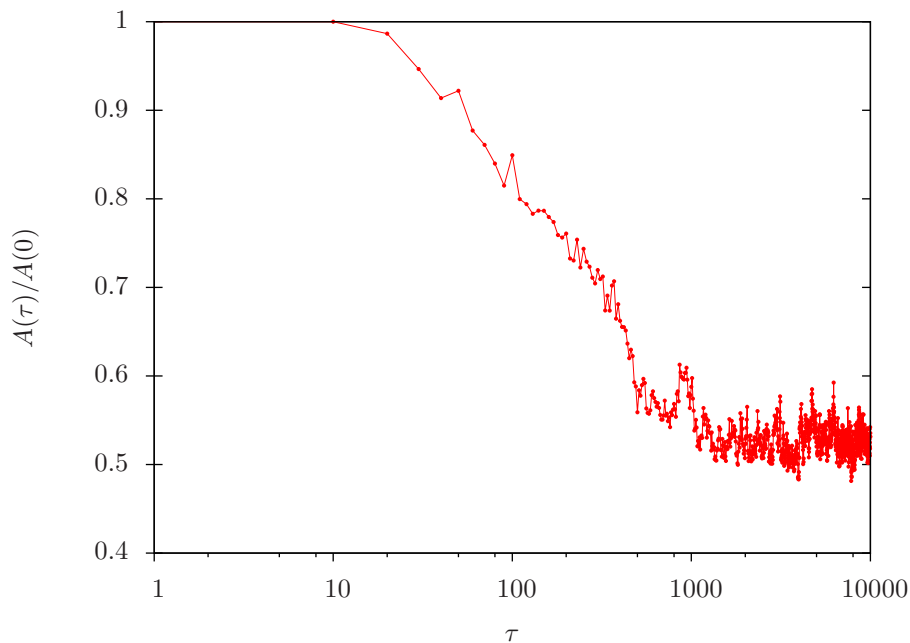}
\end{center}
\caption{ 
    { \bf Time evolution of the sum of the areas of the Voronoi polygons
    ($A(\tau)$), normalised for its value at $\tau=0$.}
A significant reduction of the domain of danger is obtained after a 
few iterations of the algorithm.
}
\label{fig_compare2}
\end{figure}

In the following, we connect the results of our analysis
with the problem of finding a good
movement rule for the selfish herd hypothesis.
To satisfy this hypothesis our algorithm must be able to generate a 
single densely packed cluster of individuals 
with a minimal area of the Voronoi polygons 
(small domains of danger) \cite{viscido2}.
Considering that we model a system with individuals characterised by
a stress zone  which is controlled by the parameter $D$, 
the ideal packed aggregation corresponds to a regular triangular lattice of side $D$.
It is clear that for  
$\Theta<220\,^{\circ}$
our algorithm achieves a perfect solution for this problem.
In contrast, if we chose $\Theta>270\,^{\circ}$ we generally obtain the 
same unsatisfactory results described
in references \cite{morton,viscido2}, 
which are characterised by a lost of cohesion.
In particular, simulations with $\Theta=360\,^{\circ}$ correspond 
to the original Hamilton's rule \cite{hamilton}.
To reinforce these considerations we calculate,
for the same parameters of Figure~\ref{fig_TransPheno1},
the mean area of the polygons at time 0 and 
for the final configuration.
Averaging over different simulations, 
we find that for $\Theta<225\,^{\circ}$ there is a reduction
of the polygons area, and thus, of the domain of danger,
of up to 75\%.
Increasing the $\Theta$ value this reduction 
diminishes and it disappears for $\Theta \ge 300\,^{\circ}$. \\

Finally, we qualitatively compare the outputs of our algorithm with
the field observations of crabs groups \cite{viscido1}. 
We consider simulations where noise
is in action and before they reach a possible
absorbing state. 
As far as data are available, we parameterise our model to 
realistic values. The parameter $D$ corresponds to the 
mean inter-individual distance after the attack 
($D=1~m$), the speed to the average crabs 
velocity during the attack  ($\epsilon=0.2~m/s$), 
and $P$ is fixed to 90 individuals.
For a square of linear size $L=14~m$ the corresponding initial density
can be put in relation with the observed one.
If the attention field
$\Theta$ is tuned to values smaller than $270\,^{\circ}$,
for a wide range of $\eta$ values,
after a rapid transient
we can observe lively shrunk configurations
qualitatively very similar to the observed real data
(see Figure~\ref{fig_compare1}).

For this set of parameters, we can state how realistic 
our simulations behave.
If we consider that fixing $\epsilon=0.2~m/s$
implies that our time unit $\tau$
corresponds to a second,
after a few seconds a significant reduction
of the domain of danger is obtained, and after
a few minutes the ideal compact configuration
is reached (see Figure~\ref{fig_compare2}).
These results can be realistically compared with the 
experimental facts.

\section*{Discussion}
\label{sec:discussion}

We introduced a simple model for exploring some aspects of order
and cohesion in swarms.
The model consists in a straightforward algorithm which states that every 
agents' motion seeks for an ideal distance from the nearest topological 
neighbour contained in a given attention field.
This approach is based on two fundamental facts present in 
nature: the existence of a limited attention field and the necessity
for a decision-making based on a minimal rule.
Despite the great simplicity of the implementation, 
varying the amplitude of the attention landscape,
the model generates a very rich behaviour: 
swarm can maintain a disordered connected shape,
it can crystallise in a six-fold ordered lattice,
it can display a low ordered phase or it can fragment and lose cohesion.
Moreover, introducing a source of noise, an 
order-disorder transition naturally appears.
These results are significant for several reasons.

First, this interaction is an ideal candidate
for solving the  ``dilemma of the selfish herd" \cite{viscido2}:  
to be able to find an easy movement rule that can produce dense aggregations.
Until now computer simulations have failed to obtain a large compact 
aggregation generated by the algorithm proposed by Hamilton:
approaching the nearest neighbour does not result in a large, dense 
group.
In contrast, the simple introduction of a reduced attention field
is able to produce the densest aggregation in 
a centrally compact swarm, with a reduction
of the domain of danger \cite{hamilton}, 
when some values of the angle 
which defines the attention field are selected.
In this perspective, the presence of a limited attention field
can be interpreted not only 
as a consequence of the constraints in the
information access or in cognitive abilities, 
but as an active regulation for reaching 
a specific collective spatial configuration.
Animals may use attention mechanisms 
to switch between processing 
few stimuli or many \cite{dukas}
in dependence of their objectives.
In our model these adjustments are obtained
modifying  the value of the parameter $\Theta$.
A single animal can switch from high values
of $\Theta$, in situations of low predation risk and foraging,
which correspond to sparse and disconnected 
configurations,
to small  values, when facing situations of danger,
which correspond to compact and dense configurations.

Second, this dynamical rule, capable of generating spatial structures 
with specific geometrical constraints, has a general interest
for collective aggregation.
In fact, 
these spatial structures
can be related to the distribution of mutual distances 
observed in surf scoters which form well-spaced groups 
of individuals on the water surface \cite{scoters}.
Moreover, our results can give some insights for 
other conventional models of collective motion.
As our study clearly evidences, local interactions with few topological 
neighbours are not just economic, but the more efficient way of 
granting the highest levels of cohesion and coherence in the swarm. 
In contrast, the interaction with a larger number of individuals
results in a loss of order and cohesion.
These outcomes are supported by 
experimental evidences in 
mosquito-fishes \cite{mosquitofish}, 
where the active regulation of the distance to the single nearest neighbour
is the fundamental interaction rule, and they
are in line with the results of different models 
which show that smaller view angles allow better cohesion
and a faster dynamics towards polarisation \cite{camperi, angle}.

Third, our results could 
be transposed into practical applications for 
designing artificial swarms builded up by cooperative mobile 
robotics \cite{martinez, sperati}. 
The studied empirically-based  algorithm could be encoded
into instructions for a scheme of distributed coordination 
to guarantee collision avoidance and cohesiveness in  
groups of autonomous agents.
From our results it follows that, for some values of the 
parameters, the final swarm configuration 
maximises the coverage of a given environment, a fact 
that can be useful for communication or detective purposes \cite{robot2}. 
A specific target application could be a swarm of 
mobile robots for environmental monitoring.

\section*{Acknowledgments}
A. M. Calv\~ao acknowledges partial financial support from a CNPq fellowship. 
\bibliography{references}

\end{document}